\documentclass[a4paper,aps,prl,preprintnumbers,amsmath,amssymb,twocolumn]{revtex4}
\usepackage{bbm}
\usepackage{amsmath}
\usepackage{verbatim}
\usepackage{graphicx}
\usepackage{epsfig}

\newcommand{\tr}[1]{{\rm tr}\left[{#1}\right]}

\newcommand{\be}{\begin{equation}}
\newcommand{\ee}{\end{equation}}
\newcommand{\bea}{\begin{eqnarray}}
\newcommand{\eea}{\end{eqnarray}}

\newcommand{\h}{{\cal H}}

\newtheorem{proposition}{Proposition}

\newtheorem{lemma}{Lemma}

\newcommand{\proofend}{\hfill\fbox\\\medskip }

\newcommand{\proof}[1]{{\bf Proof:} #1 $\proofend$}

\begin{document}
\title{Extremality of Gaussian quantum states}

\author{Michael M. Wolf$^{1}$, Geza Giedke$^{1,2}$, J. Ignacio Cirac$^{1}$}
\affiliation{$^{1}$ Max-Planck-Institute for Quantum Optics,
 Hans-Kopfermann-Str.\ 1, D-85748 Garching, Germany.\\$^{2}$ Institut f\"ur Quantenelektronik, ETH Z\"urich,
Wolfgang-Pauli-Str.16, CH-8093 Z\"urich, Switzerland.}

\begin{abstract}
We investigate Gaussian quantum states in view of their
exceptional role within the space of all continuous variables
states. A general method for deriving extremality results is
provided and applied to entanglement measures, secret key
distillation and the classical capacity of Bosonic quantum
channels. We prove that for every given covariance matrix the
distillable secret key rate and the entanglement, if measured
appropriately, are minimized by Gaussian states. This result leads
to a clearer picture of the validity of frequently made Gaussian
approximations. Moreover, it implies that Gaussian encodings are
optimal for the transmission of classical information through
Bosonic channels, if the capacity is additive.
\end{abstract}


\date{\today}

\maketitle

States with a Gaussian Wigner distribution, so called {\it
Gaussian states}, appear naturally in every quantum system which
can be described or approximated by a quadratic Bosonic
Hamiltonian. They are ubiquitous in quantum optics as well as in
the description of atomic ensembles, ion traps or nano-mechanical
oscillators. Moreover, Gaussian states became the core of quantum
information theory with continuous variables.

Besides their practical relevance, Gaussian states play  an
exceptional role with respect to many of their theoretical
properties. A particular property of Gaussian states is that they
tend to be extremal within {\it all} continuous variable states if
one imposes constraints on the covariance matrix (CM). The best
known example of that kind is the extremality with respect to the
entropy: within all states having a given CM, Gaussian states
attain the maximum von Neumann entropy (cf.\cite{maxent}). Similar
extremality properties have recently been shown for the mutual
information \cite{HolWer} and conditional entropies
\cite{Fred,JWGauss}.

In this work we prove extremality results for Gaussian states with
respect to entanglement measures, secret key rates and the
classical capacity of Bosonic quantum channels. These findings are
based on a general method, which exploits the central limit
theorem as a active and local {\it Gaussification} operation.
Our main focus lies on the entanglement, which will serve as a
showcase for the general procedure. We prove that for any given CM
the entanglement, if measured in an appropriate way, is lower
bounded by that of a Gaussian state. The same result is shown to
hold true for many other quantities like the distillable
randomness and the secret key rate. This result not only
emphasizes the exceptional role of Gaussian states, it also leads
to a clearer picture of the validity of frequently made Gaussian
approximations. In practice, states deviate from exact Gaussians
and their precise nature remains mostly unknown. Nevertheless, the
CM can typically be determined, e.g., by homodyne detection, and
one is tempted to calculate the amount of entanglement, or other
quantities, from the CM under the assumption that the state is
Gaussian. The derived extremality of Gaussian states now justifies
this approach as it excludes an overestimation of the desired
quantity, even in cases where the actual state is highly
non-Gaussian. In this sense one stays on the save side when a
priori assuming the state to be Gaussian. We will see, however,
that some care is in order, since the extremality property with
respect to the entanglement turns out to depend on the chosen
entanglement measure.

Before we derive the main results  we will briefly recall the
basic notions. Consider a Bosonic system of $N$ modes
characterized by $N$ pairs of canonical operators
$(Q_1,P_1,\ldots,Q_N,P_N)=:R$ or equivalently by $N$ Bosonic
annihilation operators $a_j=(Q_j+i P_j)/\sqrt{2}$. For any density
operator $\rho$ of the system we define a vector of means with
components $d_j=\tr{\rho R_j}$, a CM
$\Gamma_{kl}=\tr{\rho\{R_k-d_k,R_l-d_l\}_+}$ and introduce a {\it
characteristic function} $\chi(\xi)=\tr{\rho\exp{[i\xi\cdot R]}}$,
$\xi\in\mathbb{R}^{2N}$. The latter is the Fourier transform of
the Wigner function and it thus completely characterizes the state
\cite{HolevoBook}. For Gaussian states the characteristic function
has the form \be \chi(\xi)=e^{i \xi\cdot d -
\frac14\xi\cdot\Gamma\xi}\;,\ee such that  they are entirely
specified by $d$ and $\Gamma$ leading to a complete description
within a finite dimensional phase space $\mathbb{R}^{2N}$. The
underlying Hilbert space ${\cal H}$ is, however, infinite
dimensional and we will denote the set of all bounded linear
operators on ${\cal H}$ by ${\cal B(H)}$. Note that the following
results also apply to finite dimensional systems by simply
embedding $\mathbb{C}^d$ into ${\cal H}$.

Our results are based on a non-commutative central limit theorem,
as discussed in \cite{Quaegebeur,GVV,CH}, and operator-topology
arguments from \cite{CH,DD}. We will first state the main
ingredient as a general Lemma and then discuss its applications to
quantum information theory. Readers who are mainly interested in
the applications may skip the proof of the Lemma.

\begin{lemma}\label{lem1} Let $f:{\cal B}({\cal H}^{\otimes N})\rightarrow \mathbb{R}$ be a continuous functional,
 which is strongly super-additive and invariant under local unitaries $f\big(U^{\otimes N}\rho U^{\dag\otimes N}\big)=f(\rho)$. Then for every
 density operator $\rho$ describing an $N$-partite system with finite
 first and second moments, we have that
 \be\label{main}f(\rho)\geq f(\rho_G)\;,\ee
 where $\rho_G$ is the Gaussian states with the same first and
 second moments as $\rho$.
\end{lemma}

Let us first remark on the requirements in Lem.\;\ref{lem1}.
\emph{Continuity} is understood in trace-norm, i.e.,
$||\rho^{(n)}-\rho||_1\rightarrow 0$ should imply
$f\big(\rho^{(n)}\big)\rightarrow f(\rho)$. In fact, lower
semi-continuity suffices, and we may restrict the domain of $f$ to
an appropriate  compact subset of density operators like those
satisfying an energy constraint. The latter typically restores
continuity for functionals which are known to be continuous in the
finite dimensional case. \emph{Strong super-additivity} means that
given a state $\rho$ acting on
$\big(\h_{A_1}\otimes\h_{A_2}\big)\otimes\big(\h_{B_1}\otimes\h_{B_2}\big)$
with restrictions $\rho_i$ to $\h_{A_i}\otimes\h_{B_i}$, then
$f\big(\rho\big)\geq f\big(\rho_1\big)+f\big(\rho_2\big)$, with
equality if $\rho=\rho_1\otimes\rho_2$. The latter is referred to
as {\it additivity} and both properties are analogously defined
for more than two parties.

\proof{The main idea of the proof is covered by the following
equation: \bea f(\rho) &=& \frac1n f\big(\rho^{\otimes
n}\big)=\frac1n f(\tilde{\rho})\label{main1}\\
&\geq&\frac1n\sum_{k=1}^n f\big(\tilde{\rho}_k\big)\ \rightarrow\
f(\rho_G)\;.\label{main2}\eea In the first line we use additivity
of $f$ and set $\tilde{\rho}=U^{\otimes N}\rho^{\otimes
n}U^{\dag\otimes N}$, where $U$ is a suitably chosen local
unitary, which acts on  $n$ copies of the state. In the second
line we first exploit strong super-additivity in order to bound
$f(\tilde{\rho})$ from below by the sum over all reduced density
operators $\tilde{\rho}_k$. Then we argue by the central limit
theorem and a special choice of $U$, that each of these reduced
states $\tilde{\rho}_k$ converges to the Gaussian $\rho_G$ in the
limit $n\rightarrow\infty$.

This idea is now made rigorous in two steps. First we prove that
each characteristic function $\tilde{\chi}_k$ converges pointwise
to the corresponding Gaussian $\chi_G$, and then we argue that
this implies trace-norm convergence on the level of density
operators. To simplify matters we will w.l.o.g. assume that $\rho$
has vanishing first moments, i.e., $d_j=0$. The general case is
then obtained by applying local displacements, which by assumption
will not change the value of $f$.

Let us begin with specifying the local unitary $U$ as a passive
symplectic operation acting on the canonical operators on site
$\alpha\in\{ 1,\ldots,N\} $ as \be \tilde{Q}_{\alpha,k} =
\sum_{l=1}^n \frac{H_{kl}}{\sqrt{n}}\;Q_{\alpha,l}\;,\quad
\tilde{P}_{\alpha,k} =
\sum_{l=1}^n \frac{H_{kl}}{\sqrt{n}}\;P_{\alpha,l}\;,\ee with  $H={\footnotesize \left(%
\begin{array}{cc}
  1 & 1 \\
  1 &-1 \\
\end{array}%
\right)}^{\otimes m}$ being a Hadamard matrix and $n=2^m$.
Physically, $U$ corresponds to an array of 50:50 beam splitters
and half-wave plates. Note that $H$ has only entries $\pm 1$, so
that we can partition the sum \be \tilde{Q}_{\alpha,k} =
\frac1{\sqrt{n}}\left(\!\!\!\!\sum_{\ \ \ l:H_{kl}=1}^{n_+}\!\!\!
Q_{\alpha,l}\ -\!\!\!\!\!\sum_{\ \ \
j:H_{kj}=-1}^{n_-}\!\!\! Q_{\alpha,j}\right)\;, \vspace{3pt}\ee\\
and similarly for $\tilde{P}_{\alpha,k}$. Here
$n_+=n-n_-$ is the number of ones in the $k$'th row of $H$. Note
that either $n_+=n$ in the first row, or $n_+=n/2$ in all other
rows.

The characteristic function $\tilde{\chi}_k$ of the reduced
density operator $\tilde{\rho}_k$ is then given by \bea
\tilde{\chi}_k (q,p)&=&\tr{\rho^{\otimes n} \exp\left( i
\sum_{\alpha=1}^N\ q_\alpha \tilde{Q}_{\alpha,k}+p_\alpha
\tilde{P}_{\alpha,k}\right) }\\
&=&
\chi\left(\frac{\xi}{\sqrt{n}}\right)^{n_+}\chi\left(\frac{-\xi}{\sqrt{n}}\right)^{n_-}\;,\label{chipm}\eea
where $\chi$ is the characteristic function of $\rho$ and
$\xi=(q,p)$.

Following \cite{CH} we introduce a function
$g:\mathbb{R}\rightarrow\mathbb{C}$ by $ g(x)=\chi(x\xi),$ which
is a {\it classical} characteristic function, i.e., the Fourier
transform of a classical probability distribution with second
moment $ \xi^T\Gamma\xi/2$. To see this note that $\chi$ is the
Fourier transform of the Wigner function and recall that every
one-dimensional marginal of a Wigner function (in particular the
one corresponding to the direction $\xi$) is an admissible
probability distribution. Characteristic functions are continuous
at the origin, satisfy $g(0)=1$, $|g(x)|\leq 1$, and in the case
of finite second moments we can expand up to second order
\cite{Moran,exp3}, such that \be g(x)= 1- \frac{\xi^T\Gamma\xi}4
\; x^2 + o(x^2).\ee Pointwise convergence
$\tilde{\chi}_k\rightarrow\chi_G$ follows then from
Eq.(\ref{chipm}) together with setting $x=1$ in \bea
\lim_{n\rightarrow\infty}\label{classicalconv}
g\Big(\frac{x}{\sqrt{n}}\Big)^{n_+}g\Big(\frac{-x}{\sqrt{n}}\Big)^{n_-}\!\!&=&\lim_{n\rightarrow\infty}\Big(1-\frac{\xi^T\Gamma\xi}{4n}
 x^2\Big)^n\ \ \\ &=& \exp\left[-\frac14\;\xi^T\Gamma\xi x^2\right]\;.\eea
For the remaining part we can combine the argumentations in
Refs.\cite{CH,DD}. In \cite{CH} it was proven that pointwise
convergence of the characteristic functions implies convergence of
the respective density operators within the weak operator
topology. The latter was, however, shown to be equivalent to the
trace norm topology on density operators in Ref.\cite{DD}.


}

A simple application of Lem.\;\ref{lem1}   is the rederivation of the
maximum entropy principle by setting $f$
 equal to minus the von Neumann entropy $S(\rho)=-\tr{\rho\log\rho}$. Similarly, in the bipartite
case with $N=N_A+N_B$ and $f(\rho)=S(\rho_A)-S(\rho)$ we recover
the recently proven extremality result for the conditional entropy
\cite{JWGauss} for which strong super-additivity is an immediate
consequence of the strong sub-additivity inequality for the
entropy.

\section{Entanglement measures}

Grouping the $N$ tensor factors in Lem.\;\ref{lem1} into $M\leq N$ parties
and exploiting that every entanglement measure is by definition
invariant under local unitaries, yields the following.

\begin{proposition}
\label{propent} Let $E$ be a continuous entanglement measure which
is strongly super-additive. Then for every density operator $\rho$
describing an $M$-partite system with finite CM (and arbitrary,
finite, number of modes per site), we have that any Gaussian state
$\rho_G$ with the same CM provides a lower bound $E(\rho_G)\leq
E(\rho)$.
\end{proposition}

Most of the entanglement theory developed so far is devoted to
bipartite systems. Whereas for pure states there is an essentially
unique measure of entanglement, the von Neumann entropy of the
reduced state,
there are various different entanglement measures for mixed
states \cite{Hor1}. Two of them have a clear operational meaning: the {\it
distillable entanglement} $E_D$ quantifies the amount of pure
state entanglement that can asymptotically be extracted by means
of local operations and classical communication (LOCC), and the
{\it entanglement of formation} $E_F$ (resp. its regularized form,
the {\it entanglement cost} $E_c$) measures the pure state
entanglement required in order to prepare the state. Among all
other measures, the {\it logarithmic negativity} $E_N$ is the most
popular one, as it is comparatively easy to calculate \cite{VW}.
 Let us now discuss the consequences of Prop.~\ref{propent} for some entanglement
measures:

 {\it (i) Distillable entanglement:}
    $E_D$ is additive (due to the asymptotic definition) and strongly
    super-additive (since restricted protocols lead to smaller
    rates). It was shown to be continuous in the interior of the set of distillable states \cite{Guifre}. As distillable
    Gaussian states are always in the interior \cite{Gaussdist}, $E_D$ fulfills all the
    requirements in the lemma and analogous reasonings hold true in the multipartite case
    where $\omega$ is replaced by any $M$-partite target state.
    Moreover, in the bipartite case Prop.~\ref{propent} leads immediately to a simple
    sufficient distillability criterion, since for Gaussian states
    it is known that $E_D>0$ is equivalent to a simple inequality for the CM \cite{Gaussdist}.

     {\it (ii) Entanglement of formation:} Continuity of $E_F$ was
    proven in \cite{Nielsen} for finite and in
    \cite{ShirokovEoF} for infinite dimensional systems with energy constraint. Super-additivity of $E_F$ is a notorious
    conjecture, which is proven to be equivalent to additivity of
    $E_F$ and to many other additivity conjectures \cite{Shor}. These are known
    for various special cases (cf.\cite{JWGauss,Shor,WJ}) but remain to be proven in general.

    {\it (iii) Squashed entanglement} is indeed strongly
    super-additive \cite{CW} and its continuity was proven (for finite
    dimensions) in Ref. \cite{Fannes}.

     {\it (iv) Logarithmic negativity:} $E_N$ is additive, but fails to be strongly super-additive. In fact,
    $E_N$ does not only fail the requirements for the
    proposition---Eq.(\ref{main}) turns out to be false in this
    case. As simple counterexample is given by the state
    $|\varphi\rangle=\sqrt{1-\lambda^2}|00\rangle+\lambda|11\rangle$with
    $\lambda=\frac1{4}$. In this case $E_N(\varphi)\simeq 0.57$ whereas
    $E_N(\varphi_G)\simeq 0.64$ \cite{JensObs}.
    Hence, despite the fact that $E_N$ can easily be calculated,
    it is not a  faithful entanglement measure in the sense
    that a Gaussian approximation of a non-Gaussian state could lead to an over-estimated
    amount of entanglement.

Finally, it is interesting to note that Gaussian states not only
give a lower but they also provide an upper bound to the
entanglement if only the CM is known. In fact, it is a simple
consequence of the maximum entropy property that for a given
energy (i.e., $\tr{\Gamma}$ fixed) the entanglement is maximized
by a Gaussian state \cite{Enk}.

\section{secret key distillation}

We will now depart from the discussion of entanglement and see how
Lem.\;\ref{lem1} can be applied to the distillation of a classical
secret key from quantum states under the assumption of collective
attacks. Consider the case where two parties share $m$ copies of a
quantum state $\rho_{AB}$ and aim at converting these into $r m$
bits of a secret key under local operations and public
communication. Allowing for the worst case scenario, in which an
eavesdropper is given the entire purifying system of
$\rho_{AB}={\rm
tr}_E\big[|\Psi_{ABE}\rangle\langle\Psi_{ABE}|\big]$, this can be
understood as a mapping $\Psi_{ABE}^{\otimes m} \rightarrow
\sigma^{\otimes r m}\otimes \rho_E$, where the secret bits
$\sigma=\frac12\big(|00\rangle\langle00|+|11\rangle\langle11|\big)$
are asymptotically uncorrelated with the state $\rho_E$ of the
eavesdropper. We call the asymptotic supremum over all achievable
rates $r$ the {\it distillable secret key}
$K^{\mathrm{coll}}(\rho_{AB})$ of the state. By the same reasoning
as for the distillable entanglement, $K^{\mathrm{coll}}$ (together
with its multipartite generalizations with
$\sigma=\frac12\big(|0\ldots0\rangle\langle0\ldots0|+|1\ldots1\rangle\langle1\ldots1|\big)$)
has the properties of being additive and strongly super-additive.
Hence, under the assumption of continuity Lem.\;\ref{lem1} implies
the following.
\begin{proposition}
\label{propkey} Consider an $M$-partite system with an
 arbitrary, finite, number of modes per site. Then for every given finite first and second moments the Gaussian state $\rho_G$
 provides a lower bound to the distillable secret key $K^{\mathrm{coll}}(\rho)\geq K^{\mathrm{coll}}(\rho_G)$.
\end{proposition}

\section{Channel capacities}

Let us finally apply Lem.\;\ref{lem1} to the task of transmitting
classical information through a Bosonic Gaussian quantum channel
$T$ \cite{HolWer,JWGauss}. The latter may describe optical fibres,
harmonic chains or any other Bosonic system for which we can
describe the evolution in terms of a quadratic Hamiltonian $H$
acting on system plus environment \be T(\rho)={\rm
tr}_{\text{env}}\Big[V\big( \rho\otimes
|\phi\rangle\langle\phi|_{\text{env}}\big) V^\dag\Big],\quad
V=e^{iHt}\;. \ee The classical capacity $C$ of a quantum channel
is the asymptotically achievable number of classical bits that can
be reliably transmitted from a sender to a receiver per use of the
channel. To make this a reasonable concept in the infinite
dimensional setting, one imposes an energy constraint to the
encoding. That is, any allowed set of input states $\rho_i$ with
respective probabilities $p_i$ is such that the average state
$\overline{\rho}=\sum_i p_i\rho_i$ satisfies an energy constraint
$\overline{\rho}\in {\cal
K}=\{\rho|\sum_j\tr{(Q_j^2+P_j^2)\rho}\leq\kappa\}$. Under this
constraint the capacity $C(T,{\cal K})$ of the channel is  \bea
C_1(T,{\cal K}) &=& \sup_{\{p_i,\rho_i\}}
\left[S\big(T(\overline{\rho})\big)-\sum_i
p_i S\Big(T(\rho_i)\Big)\right],\\
C(T,{\cal K}) &=& \lim_{n\rightarrow\infty} \frac1n
C_1\big(T^{\otimes n},{\cal K}^{\otimes n}\big)\;. \eea Consider
now a fixed state $\overline{\rho}$ and define $\rho=V\big(
\overline{\rho}\otimes |\phi\rangle\langle\phi|_{\text{env}}\big)
V^\dag$. Then we can write \be C_1(T,\overline{\rho}) =
S\big(T(\overline{\rho})\big)-E_F\big(\rho\big)\;.\label{CEF}\ee
If the notorious additivity conjecture is true \cite{Shor}, then
not only $E_F$ satisfies the requirements of Lem.\;\ref{lem1} but
also $C_1=C$, i.e., the supremum over all $\overline{\rho}\in{\cal
K}$ in Eq.(\ref{CEF}) gives the capacity of $T$. By
Prop.~\ref{propent} together with the maximum entropy property of
Gaussian states we have then, however, that
$C(T,\overline{\rho})\leq C(T,\overline{\rho}_G)$ as both terms in
Eq.(\ref{CEF}) become extremal for the Gaussian state
$\overline{\rho}_G$, which has the same CM as $\overline{\rho}$.
Since $\overline{\rho}\in{\cal K}$ iff $\overline{\rho}_G\in{\cal
K}$ this shows:
\begin{proposition}
Consider a Bosonic Gaussian channel acting on a finite number of
modes. Then there is an optimal encoding, which achieves the
classical capacity with a Gaussian $\overline{\rho}$, if the
capacity is additive.
\end{proposition}
For single mode channels for which the optimal $\rho$ is a
symmetric two-mode Gaussian state we then even know an optimal
ensemble $\{p_i,\rho_i\}$ (which is continuous in this case). For
such two-mode states it has been shown \cite{GezaEoF} that
$E_F(\rho)$ equals the so-called {\it Gaussian entanglement of
formation} \cite{GEOF}, which in turn implies that the optimal
ensemble consists of coherent states which are distributed in
phase space according to an appropriate Gaussian
distribution.\vspace*{5pt}

There are certainly many other applications of Lem.\;\ref{lem1} proving
extremality of Gaussian states in various contexts. Via a
state-channel duality one might also apply similar techniques to
channels and operations instead of states (cf.\cite{Quaegebeur}).
In fact, recently Gaussian operations turned out to be optimal for
certain tasks concerning classical teleportation and cloning of
coherent states \cite{Gaussop}. Moreover, it is straight forward
to translate the results from the Bosonic to the Fermionic world,
i.e, in other words to work with the CAR instead of the CCR
algebra. In this case a central limit
 theorem has been proven in \cite{HudCAR}, and in \cite{GVV} the case of normal fluctuations on an
 arbitrary observable algebra was studied.

\section{Acknowledgements}

The authors are grateful to the Benasque Center for Science, where
parts of this work were developed. M.M.W. thanks R.F. Werner and
D.P. Garcia for interesting discussions.


\begin{thebibliography}{99}
\bibitem{exp3} If in addition a finite third moment exists, then
by the Berry-Esseen theorem the convergence in
Eq.(\ref{classicalconv}) is known to be uniform with convergence
rate of the order of $n^{-1/2}$.
\end{thebibliography}
\end{document}